\documentclass[conference]{IEEEtran}
\IEEEoverridecommandlockouts

\usepackage{graphicx}
\usepackage{amsmath}
\usepackage[utf8]{inputenc}
\usepackage{color}
\usepackage{graphicx}
\usepackage{rotating}
\usepackage{tikz}
\usepackage{amssymb}
\usepackage{epsfig}
\usepackage{epstopdf}
\usepackage{algorithm}
\usepackage{algorithmic}
\usepackage{amsmath}
\usepackage{amsfonts}
\usepackage{dsfont}
\usepackage{url}
\usepackage{chemarr}
\usepackage{chemarrow}
\usepackage{bbold}
\usepackage[version=4]{mhchem}

\usepackage[margin=0.65 in]{geometry}
\usepackage{mathabx}
\usepackage{tabularx,booktabs}
\usepackage{booktabs}
\usepackage{caption}
\usepackage[font=small,labelfont=bf]{caption}

\def\BibTeX{{\rm B\kern-.05em{\sc i\kern-.025em b}\kern-.08em
  T\kern-.1667em\lower.7ex\hbox{E}\kern-.125emX}}

\begin{document}

\title{Modelling the Role of Inter-cellular Communication in Modulating Photosynthesis in Plants}

 \author{\IEEEauthorblockN{Hamdan Awan$^{*}$, Raviraj S. Adve$^\diamondsuit$, Nigel Wallbridge$^\ddagger$, Carrol Plummer$^\ddagger$ and Andrew W. Eckford$^\dagger$}
 \IEEEauthorblockA{ $*$ TSSG, Waterford Institute of Technology, Ireland \\
 $^\dagger$Dept. of EECS, York University, Toronto, Ontario, Canada\\
 $^\diamondsuit$The Edward S. Rogers Sr. Dept. of ECE, University of Toronto, Ontario, Canada \\
 $^\ddagger$Vivent SaRL, Crans-pr\`es-C\'eligny, Switzerland
 }
}

\maketitle

\begin{abstract}

In this paper we show how inter-cellular molecular communication may change the overall levels of photosynthesis in plants. Individual plant cells respond to external stimuli, such as illumination levels, to regulate their photosynthetic output. Here, we present a mathematical model which shows that by sharing information internally using molecular communication, plants may increase overall photosynthate production. Numerical results show that higher mutual information between cells corresponds to an increase in overall photosynthesis by as much as 25 per cent. This suggests that molecular communication plays a vital role in maximising the photosynthesis in plants and therefore suggests new routes to influence plant development in agriculture and elsewhere.
\end{abstract}

\IEEEpeerreviewmaketitle

\section{Introduction and Related Work}
\label{sec:intro}

The diverse range of internal and external signals used by plants for regulating their functions is opening new lines of research using communication engineering theory. Cells in different plants respond to a number of external stimuli by generating information carrying \textcolor{black}{electro-chemical} signals, which propagate from one cell to another \cite{vodeneev2018parameters,surova2016variation}. This communication is used by cells to regulate various functions, such as photosynthesis. Plants cells can sense different levels of illumination (e.g. sunlight) and change their levels of photosynthesis \cite{scialdone2015plants,tedone2018plant}. Information is transmitted by different types of electro-chemical signals, such as mechanosensitive signals or \textcolor{black}{action potential} signals \cite{awan2019communication} or similar other signals. Such signals enable the transmitter cell to communicate with the neighbouring cells resulting in coherent behaviour over a region of the plant. 

In our previous works \cite{awan2019communication,awan2019acess,awan2018characterizing} we studied the generation and information theoretic properties of different types of signals in plants using molecular communication theory. The 
signalling molecules released by the transmitter cell propagate using mechanisms such as diffusion \cite{awan2016demodulation,riaz2018using} or active transport. To decode the transmitted information, signalling molecules undergo different types of interactions at the receiver such as ligand-receptor binding \cite{awan2019molecular}. \textcolor{black}{ However, in our previous works we have not studied the applications of these information carrying signals in different functionalities of plants. One such application of communication between plant cells is the regulation of the process of photosynthesis.}

\textcolor{black}{In plants, photosynthesis serves as the main mechanism for the growth by generating \textcolor{black}{photosynthate } (in presence of sunlight) during the day time.} This photosynthate is stored in starch granules which act as the reserve that degrades linearly during the night time producing sugar \cite{scialdone2015plants}. The reaction cycle of photosynthesis process consist of a large complex (reaction center) which consists of several sub-units converting light (from external stimulus) to energy. For different lengths of days the plants regulate the  production and degradation of photosynthate. This means that for longer days, a slower production rate exists for a much faster degradation rate. Similarly for shorter days, the production rate is much faster as compared to the degradation rate in the night \cite{sulpice2014arabidopsis}. 

This pattern of photosynthate production and degradation plays a key role in various other functionalities of a plant. When this process is disturbed by artificial or environmental factors (such as change in day length) the plants exhibit reduction in normal growth. However, it is observed that over time the plant can adjust the production-degradation rate in response to the environmental changes \cite{gibon2004robot}. Similarly the change in the intensity of light also results in the change in the production/degradation rate of the photosynthate produced by the plants \cite{scialdone2013arabidopsis}. This change (regulation or adjustment) in the output of photosynthesis relies on communication between plant cells. \textcolor{black}{ In literature there lacks a mathematical model for this communication between plant cells which can regulate photosynthate production (from sunlight as external stimulus).} In this paper we aim to address this gap by using the concepts of molecular communication theory. We show that by changing the parameters of molecular communication system, the plants may regulate the amount of \textcolor{black}{photosynthate} produced based on the requirements and environmental factors.

 To be precise this paper makes following two contributions: (a) To 
mathematically model the signalling mechanism in plants using molecular communication
especially in the context of regulation of photosynthesis. (b) Study the impact and relation of the mutual information between the transmitter and receiver of molecular communication system with the process of photosynthesis.

The remainder of this paper is organized as follows. \textcolor{black}{Section II describes the end to end molecular communication system model including the impact on photosynthesis and study of mutual information.  Next in Section III we present the numerical results followed by conclusion in Section IV.}

\section{Molecular Communication System Model}

The basic system model considered in this work is shown in Figure 1. We consider a system where an external stimulus (sunlight) generates an \textcolor{black}{electro-chemical} signal in the transmitter cell. \textcolor{black}{The generation of \textcolor{black}{electro-chemical} signals of different types such as \textcolor{black}{action potential} signals or mechanosensitive signals is discussed in detail in previous works  \cite{awan2019communication,awan2020communication}. Let $E_R$ represents the resting potential of the transmitter cell, then the change in membrane potential $E_m$ as a function of the changes in concentration of ions in cell membrane due to an external stimulus (i.e. sunlight) is given as:
\begin{equation}
\frac{dE_m}{dt} = \frac{1}{C} F \sum_{i}^{} z_i h_i , i  \in \textcolor{black}{({Ca^{+2}, Cl^-, K})}
\label{eq:1a}
\end{equation}
Where $F$ is Faraday's constant, $C$ is cell membrane capacitance, $z_i$ is ion $i$ charge and $h_i$ represents the ion flow in membrane as a result of change in concentrations depending on  ion-channel opening state probability $p_o$:
\begin{equation}
\frac{dp_o}{dt} = k_1 (1- p_o) - k_2 (p_o)
\end{equation}
\textcolor{black}{Note that the reaction rate constants $k_1$ (channel opening)  and $k_2$ (channel closing) depend on the membrane potential crossing a specific threshold value, therefore resulting in the generation of the electro-chemical signal $E_m$.} }The input of the molecular communication system $U(t)$ i.e. the number of signalling molecules emitted by the transmitter cell is related to the membrane potential as $U(t) \; \propto \; E_m$. This relation means that the electro-chemical signal triggers the release of proportional number of input signalling molecules which results in the inter-cellular information transfer through the diffusion of molecules. These signalling molecules propagate towards the receiver where they react with a molecular circuit to produce output molecules. 

\textcolor{black}{Next we will  present a mathematical model depicting the role of molecular communication in controlling the number of output molecules which then regulate the amount of photosynthate produced by each cell.} Furthermore, by numerical results we show that there exists an intuitive correlation between the rate of photosynthesis and mutual information of the system.

\begin{figure}
 \begin{center}
\includegraphics[trim=0cm 0cm 0cm 0cm ,clip=true, width=1\columnwidth]{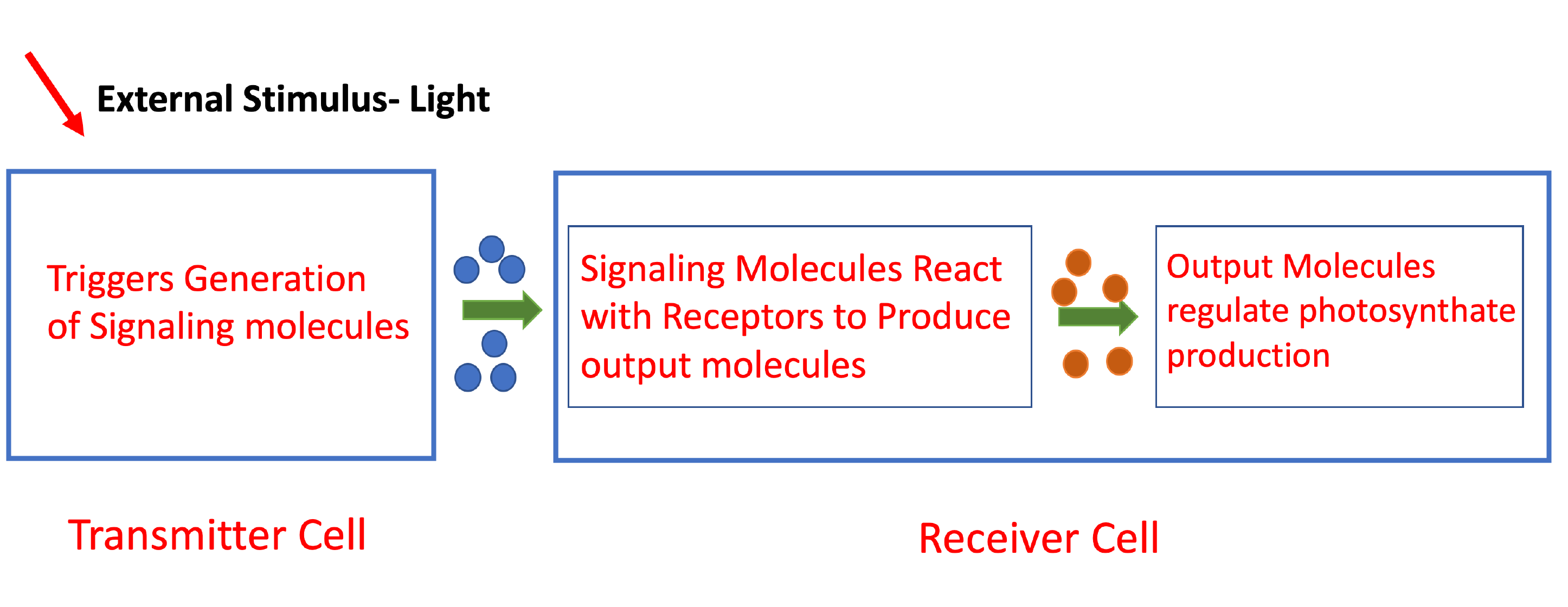}
 \caption{System Model- Transmitter-(Receiver+Photosynthesis)}
\label{system series}
 \end{center}
 \end{figure}

 \begin{figure}
 \begin{center}
\includegraphics[trim=0cm 0cm 0cm 0cm ,clip=true, width=1\columnwidth]{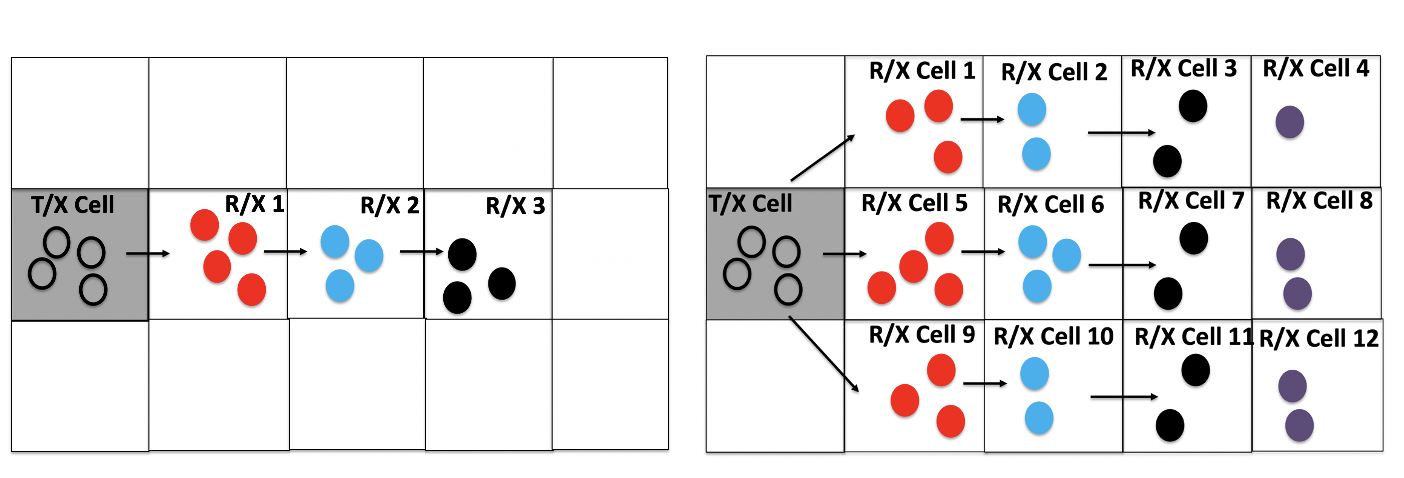}
 \caption{Voxel Model for Diffusion (T/X = Transmitter R/X= Receiver)}
\label{1c}
 \end{center}
 \end{figure}
 
\subsection{Transmitter and Propagation Model}
\label{transmitter}

We consider the system model where signalling molecules emitted by the transmitter diffuse towards the receiver cells in \textcolor{black}{ three different configurations i.e. series, parallel (for simplicity of analysis) and hybrid/mixed (more realistic) configurations.} We consider a voxel based setting to represent transmitter and receiver cells within the model. Figure \ref{1c} shows examples of voxel model for series and parallel configurations of cells. For this work we assume the transmitter and each receiver occupy a single voxel. However, it can be generalized to the case where a transmitter or a receiver occupies multiple voxels. 

The overall medium is divided into $M_x \times M_y \times M_z$ cubic voxels (each having volume $ \Delta^3$). The medium is assumed as a  3-d space of dimension $\ell_X \times \ell_Y \times \ell_Z$ such that each dimension is an integral multiple of length $ \Delta$ i.e. $\ell_X = M_x \Delta$, $\ell_Y = M_y\Delta$ and $\ell_Z = M_z \Delta $. Figure \ref{1c} shows series and parallel receiver configurations with $M_x$ = 4 and $M_y$ = $M_z$ = 1 and $M_x$ = $M_y$ = 3 and $M_z$ = 1, respectively. The voxel locations $T/X$ and $R/X_i$ represent the transmitter and receivers. \textcolor{black}{We assume a homogeneous medium with a diffusion rate $d = \frac{D}{\Delta^2}$ and diffusion coefficient $D$. The probability that a single molecule is diffused over time $\delta t$ is obtained as a product of diffusion rate $d$ with time $\delta t$ i.e. $d \delta t$.}  Let $n_{P,i}$ represents the number of molecules $P$ in any generic transmitter or receiver voxel $i$, then the state of diffusion sub-system $n_P (t)$ is given as:
\begin{equation}
n_P (t) = [n_{P,1}(t),n_{P,2}(t),n_{P,3}(t) ,n_{P,4}(t)]^T 
\label{1q:a}
\end{equation}

\textcolor{black}{The state of diffusion sub-system can change when a molecule diffuses (or jumps) from Voxel 1 to a neighboring Voxel 2 at a diffusion rate $dn_{P,1}$. This event causes $n_{P,1}$ to decrease by 1 and $n_{P,2}$ to increase by 1.  We can indicate this change by using jump vector $q_{d,1} (t) = [-1 ,1, 0, 0]^T$ and   $n_P (t)+q_{d,1}$ will be the updated state of system. The corresponding jump rate function is $W_{d,1}(n_P (t))= dn_{P,1}$ which specifies the event rate.} Combining the jump vectors and jump rate functions of all the possible diffusion  events we obtain a matrix $H$ which is employed to derive a stochastic differential equation to model the diffusion events as in \cite{awan2019molecular}: 
\begin{align}
\dot{n}_P(t) & = H n_P(t) + \sum_{j = 1}^{J_d} q_{d,j} \sqrt{W_{d,j}( n_P (t))} \gamma_j  + {\mathds 1}_T U(t)
\label{eqn:sde:do} 
\end{align}
where diffusion matrix $H$ for the entire system is given as:
\begin{align}
H =  
\left[ \begin{array}{ccccc}
-d & d & 0 & 0  \\
d & -2d & d & 0  \\
0 & d & -2d & d   \\
0 & 0 & d & -d  \\
\end{array} \right] 
\label{eqn:H} 
\end{align}
The Eq. \eqref{eqn:sde:do} is a form of a chemical Langevin equation where $\gamma_j$ is continuous-time Gaussian white noise. In Eq. \eqref{eqn:sde:do}  the first term describes the deterministic dynamics. Since all the jump rates corresponding to each diffusion event are linear, we can write this term as the product of diffusion matrix $H$ and the state vector $n_P(t)$.
Similarly, the stochastic dynamics are described by the second term. The last term models the input rate $U(t)$  of the signalling molecules emitted by the transmitter which is dependent on external stimulus (i.e. sunlight).
    

\subsection{Reaction Model}
\label{reaction}
Similar to the diffusion model, we use the stochastic differential equation (SDE) to model the dynamics of receiver reactions. For this work we assume the ligand-binding type reaction at the receiver i.e. the signalling molecules react reversibly with receptors and produce output molecules:
\begin{align}
P &  \rightarrow X 		& \left[ \begin{array}{cc} -1 & 1 \end{array} \right]^T&, k_+ n_{P,R}  \label{cr:rc1}  \\
X & \rightarrow P		& \left[ \begin{array}{cc} 1 & -1 \end{array} \right]^T&, k_- n_X       \label{cr:rc2} .
\end{align}
where $n_{P,R}$ and $n_X$ represent the number of molecules in the receiver and output molecules respectively. \textcolor{black}{The terms $k_+$ and $k_-$ are the reaction rate constants for both reactions.  For example in reaction \eqref{cr:rc1} the signaling molecules react at rate ${k}_{+} n_{L,R}$ to produce output molecules. The reaction jump vectors for each reaction indicate the change in number of signaling and output molecules.  Next we model the reaction  only system using stochastic differential equations for different receiver reactions with input $ n_{P,R} $  i.e the number of signaling molecules and the output i.e.  the number of output molecules $ n_{X} $. The state vector and SDE for reaction only system are:
\begin{align}
 \tilde{n}_R(t) 
 & =  \left[ \begin{array}{c|c}
 n_{P,R}(t) & n_X(t)  
\end{array} \right]^T 
\end{align} 
\begin{equation}
\normalsize
\dot{\tilde{n}}_R(t)  = R \tilde{n}_R(t) + \sum_{j = 1}^{2} q_{r,j} \sqrt{W_{r,j}(\langle \tilde{n}_R(t) \rangle)} \gamma_j + {\mathds 1}_T U(t) 
\label{eqn:sde:ro11} 
\end{equation}
Like the modeling of the diffusion-only case, we can use jump vectors $q_{r,j}$ and jump rates $W_{r,j}$ to model the reactions. $\gamma_j$ represents the continuous time white noise. } For the ease of mathematical modelling we define a  2$\times$2 block matrix $R$ in Table 1 with entries depending on the respective reactions of signaling molecules in the receiver.  


\begin{table}
\centering
\caption{$R$ Matrix for different receiver circuits}
\begin{tabular}{|c|c|}
\hline
\multicolumn{1}{|c|}{Receiver }	&	\multicolumn{1}{|c|}{R Matrix}	\\
\hline
Reactions 6 and 7 &    $ \begin{bmatrix}  -k_{+} & k_{-} \\ k_{+} & -k_{-} \end{bmatrix}$
\\ \hline
\end{tabular}
\label{table:1a}
\end{table}

\subsection{SDE for End to End Model}
\label{complete}



In this section we derive the SDE for the end to end system (accounting for both diffusion and reaction events) by interconnection of the SDE derived separately for diffusion and reaction events separately in Sections IIA and IIB. This interconnection between the diffusion-only subsystem and the reaction-only subsystem is the number of signaling molecules in the receiver voxel  $n_{P,R}(t)$, which is common to both of them i.e. it appears in the state vectors $n_P(t)$ and $\tilde{n}_R(t)$ of both diffusion only and receiver only modules respectively as shown in Eqs. (3) and (8).  

\textcolor{black}{For the end to end system the dynamics of $n_{P,R}(t)$ is obtained by combining the impact of both diffusion and reactions on the dynamics of signaling molecules in system.} Let $n(t)$ be the state of the complete system  given by: 
\begin{align}
n(t) = 
 & \left[ \begin{array}{c|c}
 n_{P}(t)^T & n_X(t)  
\end{array} \right]^T
\label{eqn:state} 
\end{align}
where $n_{P}(t)$ and $n_{X}(t)$ represent the number of input signalling molecules $P$ and number of output molecules $X$ respectively. We modify the jump vectors from the diffusion-only subsystem and the reaction-only subsystem, to obtain the jump vectors for the complete model.  We use $q_j$ and $W_j(n(t))$ to denote the jump vectors and jump rates of the combined model. The SDE for the complete system is then obtained by combining Eqs. (4) and (9) as:
\begin{align}
\dot{n}(t) & = A n(t) + \sum_{i = 1}^{J} q_j \sqrt{W_j(n(t))} \gamma_j + {\mathds 1}_T U(t) 
\label{eqn:mas11a}
\end{align} 
where $J$ is the sum of all events (i.e. reactions and diffusion). We define matrix $A$ by $A n(t) = \sum_{i = 1}^{J} q_j W_j(n(t))$ which is obtained as:
\begin{align}
A = 
 & \left[ \begin{array}{c|c}
H + {\mathds 1}_R^T {\mathds 1}_R R_{11}  &   {\mathds 1}_R R_{12}  \\ \hline 
R_{21}  {\mathds 1}_R^T & R_{22} 
\end{array} \right]
\label{eqn:A} 
\end{align}
where diffusion matrix $H$ (given in \eqref{eqn:H}) is the infinitesimal generator of a Markov chain governing the diffusion of signalling molecules. The $R_{ii}$ terms account for the reaction events in system as shown in Table 1. Next we perform Laplace transform of \eqref{eqn:mas11a} to obtain $N_X(s)$ i.e. number of output molecules in the system as:
\begin{align}
  \langle N_X(s) \rangle = {\mathds 1}_X  \langle N(s)\rangle =
 \underbrace{ {\mathds 1}_X  (sI - { A})^{-1} {\mathds 1}_T }_{\Psi(s)}   U(s)
\label{eqn:mas11}
\end{align}
Next we show that how these output molecules are responsible for regulating the photosynthate production of the plant cells.

\subsection{Impact on Photosynthesis}

Photosynthesis is a process where water ($H_2 O$) is combined with carbon dioxide ($CO_2$) to produce glucose ($C_6H_{12}O_6$) (i.e. photosynthate $p_h$) and molecular oxygen ($O_2$). This process relies on the energy from the sunlight (i.e. external stimulus) which is stored as chemical potential energy in the output molecules $n_X$ of the molecular communication system. The overall balanced equation of the reaction is:
\begin{equation}
\ce{6CO2 + 6H2O + n_X (Energy) -> C6H12O6 + 6O2}
\end{equation}
We describe the dynamics of energy required for photosynthesis in plant cells, which shows that the number of output molecules $n_X$ may vary depending on the parameters of the molecular circuit. For example by changing the reaction rate constants of molecular receiver circuit, we can increase or decrease the number of output molecules and therefore the photosynthate $p_h$ produced (as shown in the system model in Fig.1), which is used for maintenance of different functionalities of plants. To  model the photosynthesis process as a time-dependent function (where photosynthate is produced by each gram of photo-synthetically active biomass) we use:
\begin{equation}
  p_h=  L C p_{h}^{max} min (n_X(t)) = L C p_{h}^{max} n_X^{ph}(t)
  \label{new}
\end{equation}
where the term  $p_h$ represents the number of photosynthate produced. $p_{h}^{max}$ is the maximum rate of the photosynthesis. \textcolor{black}{ The term $n_X^{ph}(t)$ or $min$ $(n_X(t))$ represents the minimum number of output molecules required to sustain the maximum rate of photosynthesis which is obtained in Section IIC.} $L $ is a binary function separating day and night. We can modify $L$ as a continuous time-dependent function such that $L \in  [0, 1]$. $C \in  [0, 1]$ is a positive function representing the limiting coefficient to photosynthesis. In this model $C$ proposes a limiting function depending on a range of factors as shown in following:
  \begin{equation}
  C= \lambda_c (1- \lambda_c) \frac{a_{max}-a}{a_{max}}
  \label{newA}
\end{equation}
where  $a$ is the starch, $\lambda_c$ represents the strength of photosynthesis limitation and $a_{max}$ is the maximum starch that can be consumed at the maximum rate $\tau_{as}^{max}$. From Eqs. \eqref{new} and \eqref{newA} we learn that the output of photosynthesis is correlated strictly with availability of the number of output molecules to be above a certain threshold. Note that regardless of the duration of night, the photosynthate $p_h$ degrades linearly at a rate $\tau_{as}$ and is almost consumed before the day time \cite{scialdone2015plants, tedone2018plant}.

\subsection{Mutual Information}
In this section we will calculate the mutual information of the inter-cellular molecular communication system to quantify its impact on the process of photosynthesis. The input of the end to end system is the input rate of signalling molecules $U(t)$ whereas the  output of the system is  $n_{X}(t)$ i.e. number of output molecules controlling the process of photosynthesis as shown in Eq. (15). The expression for the mutual information $I(n_{X},U)$ between two Gaussian distributed random processes $U(t)$ and $n_{X}(t)$  is given as:
\begin{align}
I(n_{X},U) = \frac{1}{2} \int \log \left( 1+\frac{ | \Psi(\omega) |^2}{\Phi_{\eta}(\omega)} \Phi_u(\omega) \right) d\omega
\label{eqn:mi1}
\end{align}

\begin{table}[t]
\centering
\caption{Parameters and their default values.}
\begin{tabular}{|c|c|}
\hline
\multicolumn{1}{|c|}{Symbols}	&	\multicolumn{1}{|c|}{Notation and Value}	
\\ \hline
$E_R$ &     Resting Potential of cell membrane  = -150-170 Milli volts.
\\ \hline
$F$ &    Faraday's constant = $9.65 \times 10^4 C/mol$ 
\\ \hline
$C$ &     Membrane capacity = $10^{-6} F cm^-2$  
\\ \hline
$P_m$  &     Permeability per unit area  = $10^{-6}$ M cm $s^{-1}$
\\ \hline
$\gamma$ &  association-disassociation rate constants ratio = $9.9 \times  10^-5 M$ 
\\ \hline
 $\phi_{i}$ & Probability ion link to channel inside $c_{in}$ / ($c_{in}$ + $\gamma$)
\\ \hline
 $\phi_{o}$ & Probability ion link to channel outside $c_{out}$ / ($c_{out}$ + $\gamma$)
 \\ \hline
 $\eta_i$ & Probability ion not linked to channel inside = 1- $\phi_{i}$ 
\\ \hline
 $\eta_o$ & Probability ion not linked to channel outside = 1- $\phi_{o}$
\\ \hline
$c_{in}$  &  1.28	
\\ \hline
$c_{out}$ &  1.15 
\\ \hline
$z$   & ion charge e.g. for calcium = +2  
\\ \hline
\end{tabular}
\label{table:2}
\end{table}

\textcolor{black}{When we assume all the chemical reactions and corresponding jump rates as linear the power spectral density of $n(t)$ is obtained by using Eq. (11). The noise in the output $n_X(t)$ is caused by the Gaussian white noise variable $\gamma_j$. Therefore, Eq.(11) models a continuous-time linear time-invariant (LTI) stochastic system subject to Gaussian input and Gaussian noise.  \textcolor{black}{The power spectral density $\Phi_{X}(\omega)$ of the signal $n_X(t)$ can be obtained from standard results on the output response of a LTI system to a stationary input as: }
\begin{align}
\Phi_{{X}}(\omega) & =  |\Psi(\omega) |^2 \Phi_e(\omega) + \Phi_{\eta}(\omega)
\label{2331a} 
\end{align}
where $\Phi_u(\omega)$ is the power spectral density of $U(t)$ and $|\Psi(\omega)|^2$ is the channel gain with $\Psi(\omega) = \Psi(s)|_{s = i\omega}$. The term $\Phi_{\eta}(\omega)$ denotes the stationary noise spectrum and is given by: 
\begin{align}
\Phi_{\eta}(\omega) & =   \sum_{j = 1}^{J_d + J_r} | {\mathds 1}_X (i \omega I - A)^{-1} q_j |^2 W_j(\langle n_{}(\infty) \rangle) 
\label{eqn:spec:noise2} 
\end{align} 
where $n_{}(t)$ denotes the state of the complete system in Eq. \eqref{eqn:state} and $\langle n_{}(\infty) \rangle$ is the mean state of system at time $\infty$ due to constant input $c$. Similarly, by using standard results on the LTI system, the cross spectral density $\Psi_{xu}(\omega)$ has the following property:
\begin{align}
|\Psi_{xu}(\omega)|^2 &= |\Psi(\omega) |^2 \Phi_u(\omega)^2 
\label{eqn:csd} 
\end{align} 
Finally, we apply the water filling solution \cite{gallager1968information} to Eq. (19) subject to a power constraint on the input $U(t)$ to maximize the mutual information of the system.}

\section{Numerical Results and Discussion}
\textcolor{black}{In this work we use the parameters for the generation of \textcolor{black}{electro-chemical} signals due to an external stimulus (i.e. sunlight) similar to \cite{awan2019communication} which are presented in Table II. } The parameters related to photosynthate production are obtained from \cite{scialdone2015plants,tedone2018plant}.  To model the diffusion of molecules the size of each voxel is assumed to be  ($\frac{1}{3}$ $\mu$m)$^{3}$ (i.e., $\Delta = \frac{1}{3}$ $\mu$m). The size of medium is assumed as 2$\mu$m $\times$ 2 $\mu$m $\times$ 1 $\mu$m leading to an array of 6 $\times$ 6 $\times$ 3 voxels. Furthermore we assume that both the transmitter and receiver occupy one voxel each. \textcolor{black}{The results presented in this section are validated by simulations similar to  previous works \cite{awan2020communication}, however they are not included in this paper due to limitation of space.}

First, in Figure 3  we show that the number of output molecules (proportional to photosynthate) tends to increase with the increase in the reaction rate constant $k_+$ in the receiver molecular circuit for all three configurations of receiver cells considered in this paper. This suggests that plants may control (increase and decrease) the production of photosynthate by inter-cellular communication in plants. 

 \begin{figure}[h]
\begin{center}
\includegraphics[trim=0cm 0cm 0cm 0cm ,clip=true, width=0.9\columnwidth]{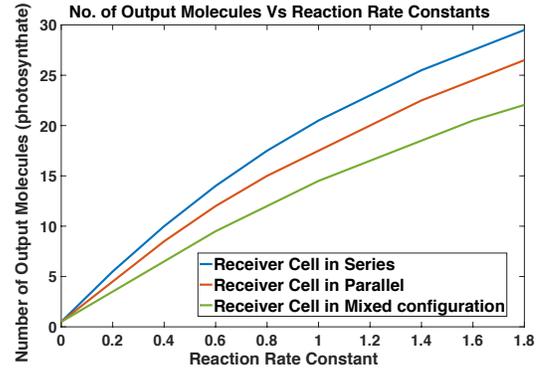}
\caption{Output Molecules vs Reaction Rate Constant $k_+$}
\end{center}
\end{figure}

\begin{figure}[h]
\begin{center}
\includegraphics[trim=0cm 0cm 0cm 0cm ,clip=true, width=0.85\columnwidth]{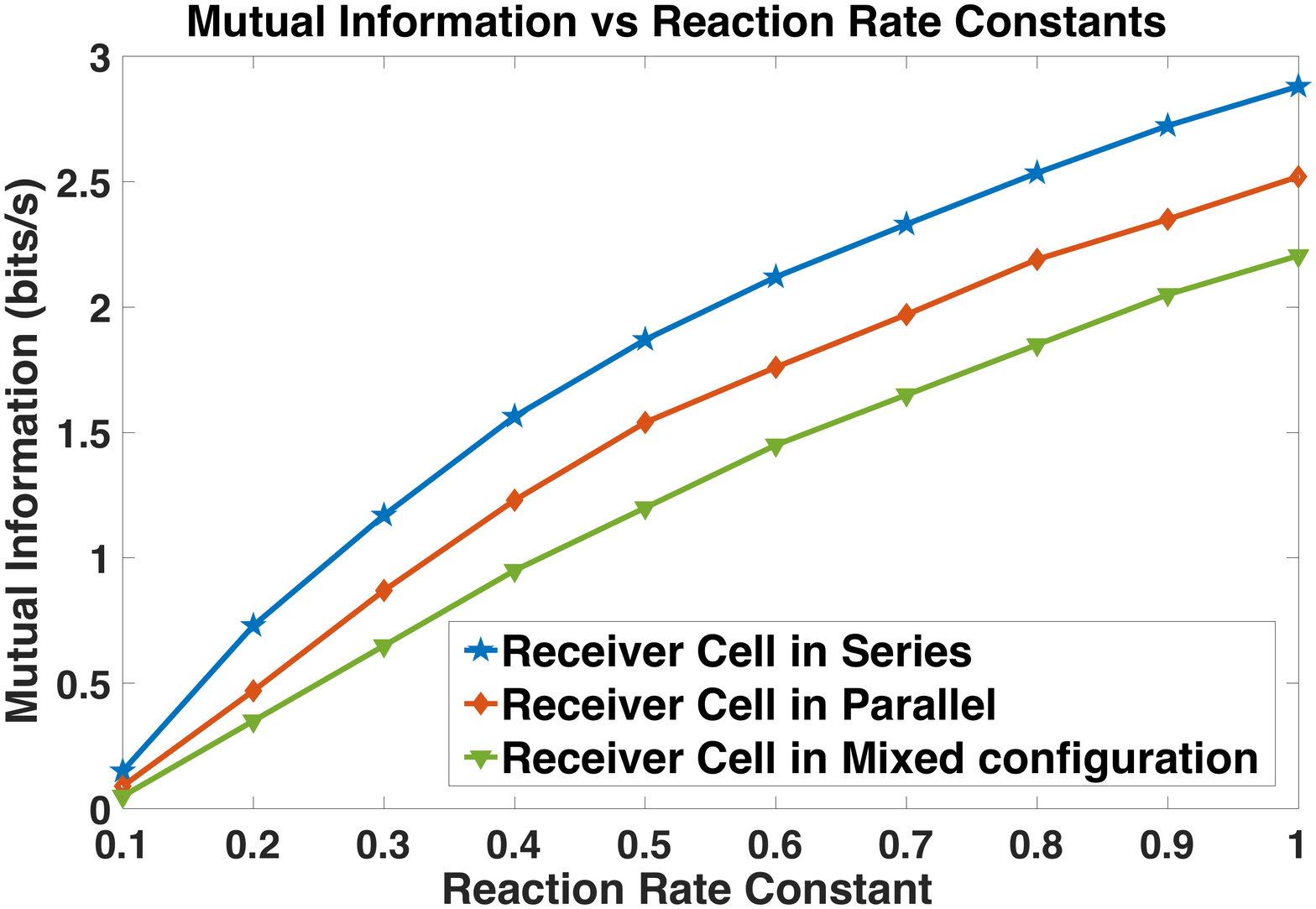}
\caption{Mutual Information vs Reaction Rate Constant $k_+$}
\end{center}
\end{figure}

\textcolor{black}{This result also provides an explanation for the self-regulatory mechanism employed by different plants depending on the individual needs, environmental factors and available resources. Next in Figure 4 we show that as we increase the reaction rate constant $k_+$, the mutual information also tends to increase for all three configurations of receiver cells. Note that  the results in both Figures 3 and 4 show that the receiver cells in series lead to a higher no. of output molecules as well as mutual information in comparison to parallel and mixed configuration. This is due to higher loss of input signalling molecules in the parallel and mixed configuration of cells. }

Next in Figure 5 we show the relation between the \textcolor{black}{photosynthate} produced over time and the potential corresponding to the external stress. \textcolor{black}{Since our model is generic and can be applied to different types of configurations we have only included the results for series configuration of cells due to limitation of space. Note that the results for parallel and mixed configurations of cells follow a similar pattern.} The result in Figure 5 shows that as the intensity of external stimulus increases the number of photosynthate produced also increases. Similarly in Figure 6 we show that mutual information corresponding to inter-cellular molecular communication also increases with the increase in external stimulus levels. 

\textcolor{black}{From these results we learn that the mutual information increases as the number of output molecules produced in the system increase. \textcolor{black}{This suggests that a higher number of output molecules and hence a higher mutual information results in an increase in the output of the photosynthesis process.}  We also learn that by modifying mutual information of molecular communication (through the system parameters) we may control (increase or decrease up to 25 per cent) the rate of photosynthesis in the receiver cells. These results suggest an important role of inter-cellular molecular communication in plants to regulate the important function of growth i.e. photosynthesis.}

 \begin{figure}
 \begin{center}
\includegraphics[trim=0cm 0cm 0cm 0cm ,clip=true, width=0.85\columnwidth]{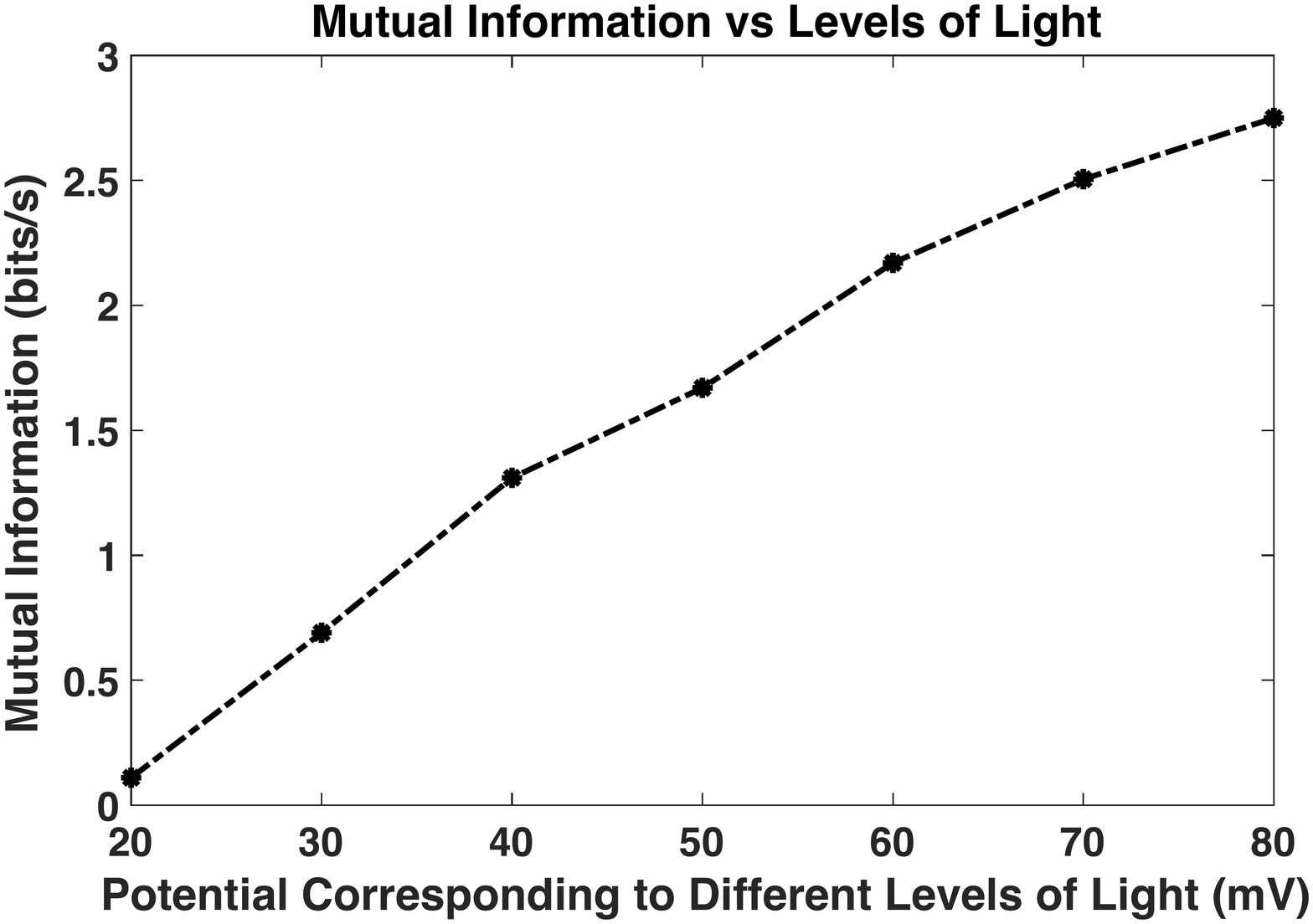}
 \caption{ Number of Photosynthate vs increasing Light Levels}
\label{s1}
 \end{center}
 \end{figure}
 
  \begin{figure}
 \begin{center}
\includegraphics[trim=0cm 0cm 0cm 0cm ,clip=true, width=0.85\columnwidth]{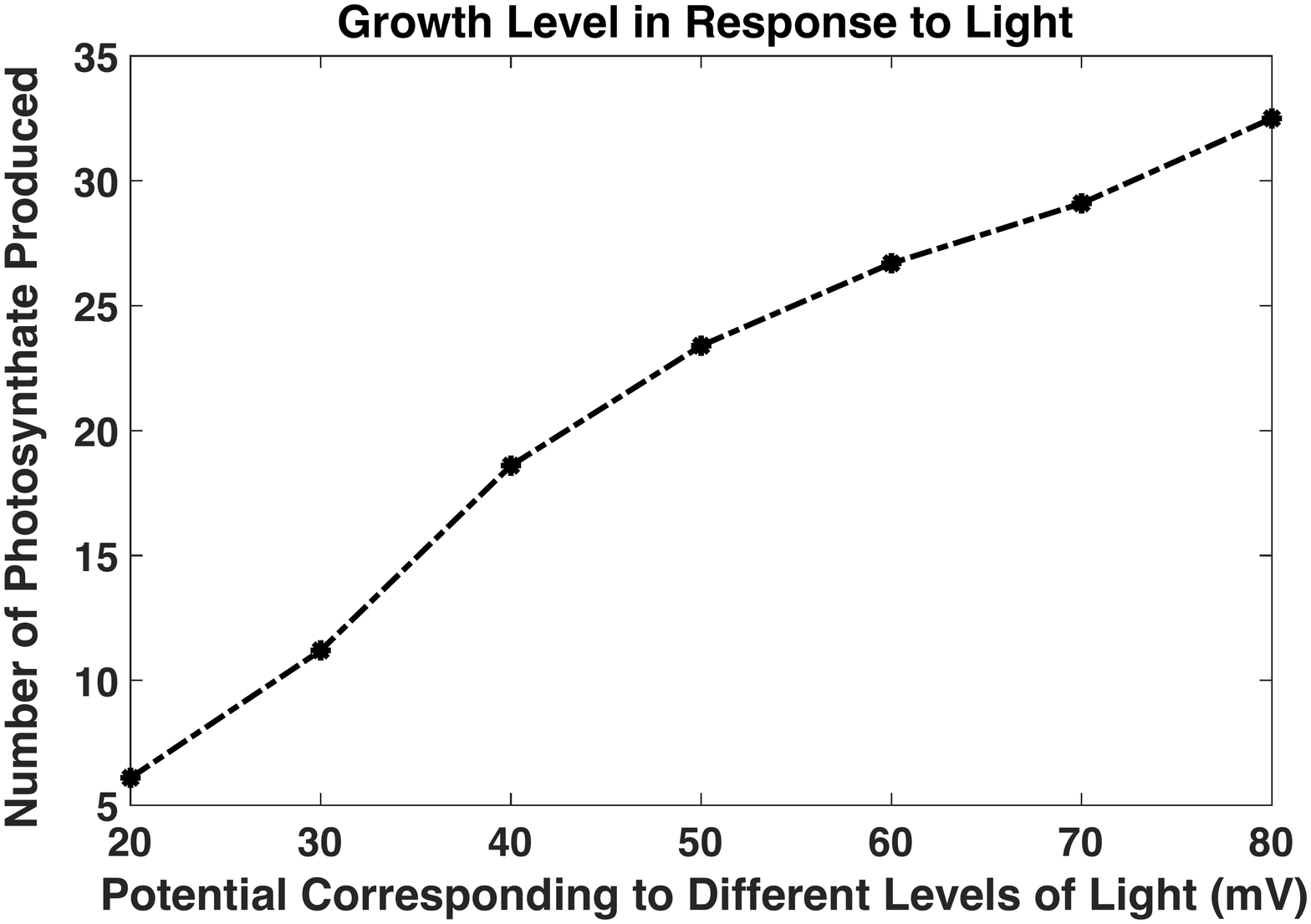}
 \caption{Mutual Information vs increasing Light Levels}
\label{s2}
 \end{center}
 \end{figure}

\section{Conclusion}

This work presents a mathematical model for regulating the output of photosynthesis in plants by using inter-cellular molecular communication. We show that both the photosynthate production and mutual information of molecular communication system tends to increase simultaneously as the external stimulus i.e. light intensity increases. Varying the parameters of molecular communication system (i.e. reaction rate constants of receiver circuit) the rate of photosynthesis may be increased or decreased. We further realise that the external stimulus (light) may also impact the photosynthate production in the neighbouring cells due to the inter-cellular molecular communication. This suggests that inter-cellular communication has a significant impact on plant productivity. \textcolor{black}{In the future work we aim to conduct experiments to validate the proposed theoretical model in this paper.}


\ifCLASSOPTIONcaptionsoff
  \newpage
\fi

\bibliographystyle{ieeetr}
\bibliography{nano2018}

\begin{thebibliography}{10}

\bibitem{vodeneev2018parameters}
V.~Vodeneev, M.~Mudrilov, E.~Akinchits, I.~Balalaeva, and V.~Sukhov,
  ``Parameters of electrical signals and photosynthetic responses induced by
  them in pea seedlings depend on the nature of stimulus,'' {\em Functional
  Plant Biology}, vol.~45, no.~2, pp.~160--170, 2018.

\bibitem{surova2016variation}
L.~Surova, O.~Sherstneva, V.~Vodeneev, and V.~Sukhov, ``Variation potential
  propagation decreases heat-related damage of pea photosystem i by 2 different
  pathways,'' {\em Plant signaling \& behavior}, vol.~11, no.~3, p.~e1145334,
  2016.

\bibitem{scialdone2015plants}
A.~Scialdone and M.~Howard, ``How plants manage food reserves at night:
  quantitative models and open questions,'' {\em Frontiers in plant science},
  vol.~6, p.~204, 2015.

\bibitem{tedone2018plant}
F.~Tedone, E.~Del~Dottore, B.~Mazzolai, and P.~Marcati, ``Plant behaviour: a
  mathematical approach for understanding intra-plant communication,'' {\em
  BioRxiv}, p.~493999, 2018.

\bibitem{awan2019communication}
H.~Awan, R.~S. Adve, N.~Wallbridge, C.~Plummer, and A.~W. Eckford,
  ``Communication and information theory of single action potential signals in
  plants,'' {\em IEEE transactions on nanobioscience}, vol.~18, no.~1,
  pp.~61--73, 2019.

\bibitem{awan2019acess}
H.~Awan, R.~S. Adve, N.~Wallbridge, C.~Plummer, and A.~W. Eckford,
  ``Information theoretic based comparative analysis of different communication
  signals in plants,'' {\em IEEE Access}, Aug 2019.

\bibitem{awan2018characterizing}
H.~Awan, R.~S. Adve, N.~Wallbridge, C.~Plummer, and A.~W. Eckford,
  ``Communication and information theory of single action potential signals in
  plants,'' {\em IEEE transactions on nanobioscience}, 2018.

\bibitem{awan2016demodulation}
H.~Awan and C.~T. Chou, ``Demodulation of reaction shift keying signals in
  molecular communication network with protein kinase receiver circuit,'' in
  {\em Wireless Communications and Networking Conference (WCNC), 2016 IEEE},
  IEEE, 2016.

\bibitem{riaz2018using}
M.~U. Riaz, H.~Awan, and C.~T. Chou, ``Using spatial partitioning to reduce
  receiver signal variance in diffusion-based molecular communication,'' in
  {\em Proceedings of the 5th ACM International Conference on Nanoscale
  Computing and Communication}, pp.~1--6, 2018.

\bibitem{awan2019molecular}
H.~Awan and C.~T. Chou, ``Molecular communications with molecular circuit-based
  transmitters and receivers,'' {\em IEEE transactions on nanobioscience},
  vol.~18, no.~2, pp.~146--155, 2019.

\bibitem{sulpice2014arabidopsis}
R.~Sulpice, A.~Flis, A.~A. Ivakov, F.~Apelt, N.~Krohn, B.~Encke, C.~Abel,
  R.~Feil, J.~E. Lunn, and M.~Stitt, ``Arabidopsis coordinates the diurnal
  regulation of carbon allocation and growth across a wide range of
  photoperiods,'' {\em Molecular Plant}, vol.~7, no.~1, pp.~137--155, 2014.

\bibitem{gibon2004robot}
Y.~Gibon~et al, ``A robot-based platform to measure multiple enzyme activities
  in arabidopsis using a set of cycling assays: comparison of changes of enzyme
  activities and transcript levels during diurnal cycles and in prolonged
  darkness,'' {\em The Plant Cell}, vol.~16, no.~12, 2004.

\bibitem{scialdone2013arabidopsis}
A.~Scialdone, S.~T. Mugford, D.~Feike, A.~Skeffington, P.~Borrill, A.~Graf,
  A.~M. Smith, and M.~Howard, ``Arabidopsis plants perform arithmetic division
  to prevent starvation at night,'' {\em Elife}, vol.~2, p.~e00669, 2013.

\bibitem{awan2020communication}
H.~Awan, K.~Zeid, R.~S. Adve, N.~Wallbridge, C.~Plummer, and A.~W. Eckford,
  ``Communication in plants: Comparison of multiple action potential and
  mechanosensitive signals with experiments,'' {\em IEEE Transactions on
  NanoBioscience}, vol.~19, no.~2, pp.~213--223, 2019.

\bibitem{gallager1968information}
R.~G. Gallager, {\em Information theory and reliable communication}, vol.~2.
\newblock Springer, 1968.

\end{thebibliography}
\begin{IEEEbiography}[{\includegraphics[width=1in,height=1.25in,clip,keepaspectratio]{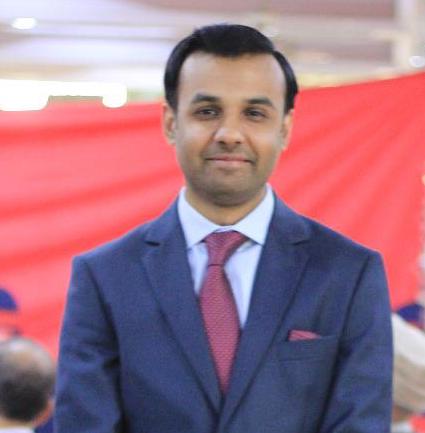}}]{Hamdan Awan}  has received the Ph.D. degree in computer science and engineering from the University of New South Wales (UNSW), Sydney, Australia, in August 2017. He stayed with York University, Canada, as a Post-Doctoral Fellow, for two years, from December 2017 to December 2019, where he worked on the DARPA’S Radio Bio Project. In December 2019, he joined the Telecommunications Systems and Software Research Group, Waterford Institute of Technology, Ireland, as an H2020 Research Fellow, where he is working on the FET-Open Gladiator Project and FET-Open Prime Project. He has so far published more than 30 research papers in highly selective IEEE Transactions and the IEEE/ACM conferences. He has also co-authored the article that received the Best Paper Award from the IEEE/ACM NanoCom Conference, in 2018. His major research interests include molecular communications, nano-networks, information theory aspects of biological communication, and computer vision. He has also been a recipient of Post Doctoral Writing Fellowship after the Ph.D. degree at the UNSW, for three months, from September to November 2017.
\end{IEEEbiography}

\begin{IEEEbiography}[{\includegraphics[width=1in,height=1.25in,clip,keepaspectratio]{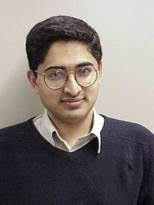}}]{Raviraj S.Adve}  was born in Bombay, India. He received his B. Tech. in Electrical Engineering from IIT, Bombay, in 1990 and his Ph.D. from Syracuse University in 1996, His thesis received the Syracuse University Outstanding Dissertation Award. Between 1997 and August 2000, he worked for Research Associates for Defense Conversion Inc. on contract with the Air Force Research Laboratory at Rome, NY. He joined the faculty at the University of Toronto in August 2000 where he is currently a Professor. Dr. Adve’s research interests include molecular communications, analysis and design techniques for cooperative and heterogeneous networks, energy harvesting networks and in signal processing techniques for radar and sonar systems. He received the 2009 Fred Nathanson Young Radar Engineer of the Year award. Dr. Adve is a Fellow of the IEEE.
\end{IEEEbiography}

\begin{IEEEbiography}[{\includegraphics[width=1in,height=1.25in,clip,keepaspectratio]{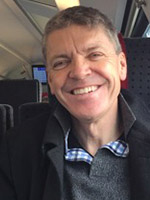}}]{Nigel Wallbridge}  is an internationally renowned serial entrepreneur with extensive experience found- ing and growing companies in both North America and Europe. A chartered engineer, he holds a doctoral degree in medical engineering from the University of Leeds in the UK and also an MBA from INSEAD, France. Apart from his current work at Vivent, he is also currently an Executive Board Member at TeleRail Networks Ltd and is on the board of a number of other technology businesses. He previously held senior managerial positions in the telecommunications industry, including CEO of Interoute, and President of Cable and Wireless Americas. He also lectured in medical engineering at the University of Leeds. 
\end{IEEEbiography}

\begin{IEEEbiography}[{\includegraphics[width=1in,height=1.25in,clip,keepaspectratio]{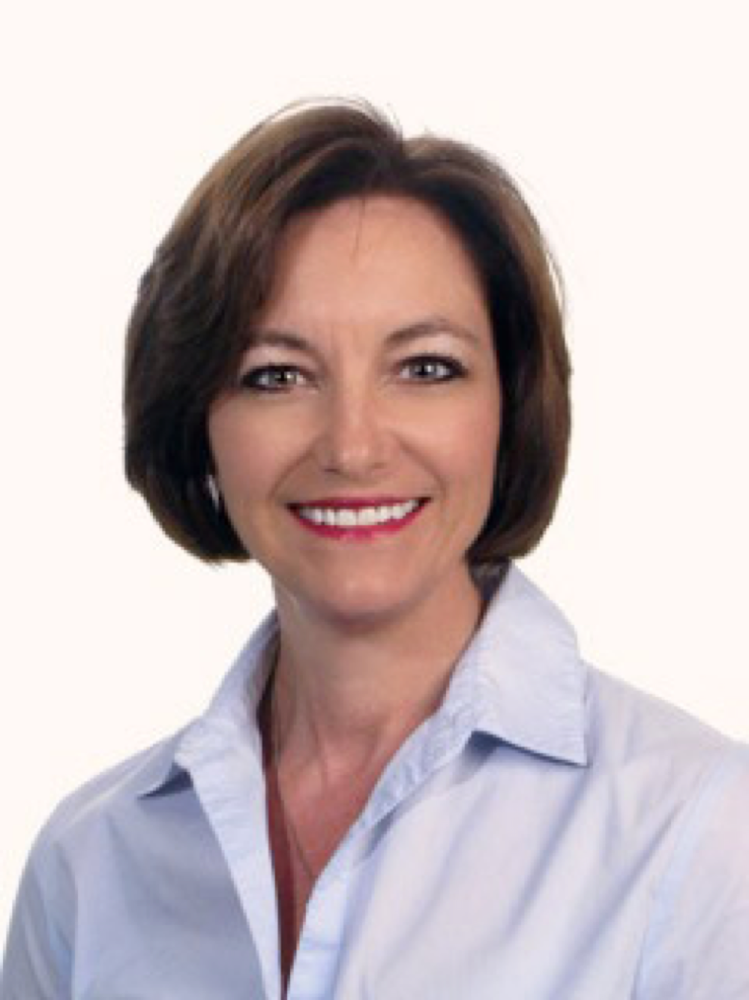}}]{Carrol Plummer} is the CEO and a co-founder of Vivent SARL, a research intensive Swiss based business focused on understanding communications net- works in biological systems. She received her BSc in Mechanical Engineering from University of Calgary in 1981, and an MBA from INSEAD, France in 1990. Her research interests include mechanical and electrical signaling in plants, bacteria and animals from both theoretical and applied perspectives, and the role signaling plays in organism wide coordinated responses such as growth or defense reactions. She is currently working on research supported by grants from DARPA and Innosuisse, with a focus on the use of machine learning to analyse electrophysiology signals. 
\end{IEEEbiography}

\begin{IEEEbiography}[{\includegraphics[width=1in,height=1.25in,clip,keepaspectratio]{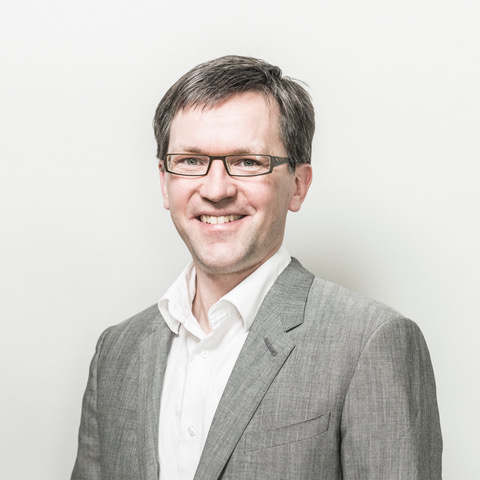}}]{Andrew W. Eckford}  is an Associate Professor in York University,  Toronto, Ontario. He received degrees from the University of Toronto in 1999 and 2004, respectively, all in Electrical Engineering. Andrew held postdoctoral fellowships at the University of Notre Dame and the University of Toronto, prior to taking up a faculty position at York in 2006. Andrew’s research interests include the application of information theory to nonconventional channels and systems, especially the use of molecular and biological means to communicate. Andrew’s research has been covered in media including The Economist and The Wall Street Journal and was a finalist for the Bell Labs Prize. Andrew is also a co-author of the textbook Molecular Communication, published by Cambridge University Press. 
\end{IEEEbiography}

\end{document}